\documentclass{ws-procs975x65}
\DeclareMathOperator{\tr}{Tr}
\begin{document}

\title{Bose-Einstein condensation on quantum graphs}

\author{J. Bolte}

\address{Department of Mathematics, Royal Holloway, University of London\\
Egham, TW20 0EX, United Kingdom\\
E-mail: jens.bolte@rhul.ac.uk}

\author{J. Kerner}

\address{Institut f\"{u}r Analysis, Dynamik und Modellierung, Universit\"{a}t 
Stuttgart,\\ 70565 Stuttgart, Germany\\
E-mail: Joachim.Kerner@mathematik.uni-stuttgart.de}

\begin{abstract}
We present results on Bose-Einstein condensation (BEC) on general compact 
quantum graphs, i.e., one-dimensional systems with a (potentially) complex 
topology. We first investigate non-interacting many-particle systems and 
provide a complete classification of systems that exhibit condensation. 
We then consider models with interactions that consist of a singular part as 
well as a hardcore part. In this way we obtain generalisations of the 
Tonks-Girardeau gas to graphs. For this we find an absence of phase transitions which then indicates 
an absence of BEC.
\end{abstract}


\bodymatter

\section{Introduction}
\label{aba:sec1}
We present results on Bose-Einstein condensation (BEC) in many-particle 
systems on compact quantum graphs.\cite{BKSingular,BKContact,BKBEC} 
Quantum graphs are models of particles moving along the edges of a metric 
graph. They hence combine the simplicity of a one-dimensional model with the 
complexity of a graph. One of the major findings about quantum graphs was that 
the one-particle spectra display correlations that are well described by 
random-matrix theory.\cite{KS97,GNUSMY06} For that reason quantum graphs have 
become popular models in quantum chaos.

Bose-Einstein condensation, on the other hand, is a quantum mechanical 
phenomenon of many-particle systems that can be well described within the 
framework of quantum statistical mechanics. Originally, this condensation was 
predicted by Einstein for a gas of non-interacting bosons in three 
dimensions.\cite{EinsteinBEC} The conclusion was that below some critical 
temperature $T_{crit}$, or above some critical particle density $\rho_{crit}$, 
all additional particles brought into the system must condense into the 
one-particle ground state. In other words, the one-particle ground state 
becomes macroscopically occupied. As recognised later,\cite{PO56} the 
macroscopic occupation of a \textit{single-particle} state is indeed the 
underlying mechanism that gives rise to the condensation, also in interacting 
many-particle system

We here introduce many-particle systems on graphs and study BEC in 
non-interacting systems as well as in systems with hardcore interactions, thus 
generalising the Tonks-Girardeau gas\cite{G60} to graphs.
\section{Preliminaries}
In this section we briefly summarise relevant concepts of BEC as well as of 
many-particle quantum graphs. For more details on BEC see 
Refs.~\citenum{PO56,CCGOR11}, on quantum graphs see 
Refs.~\citenum{KS06,Kuc04,GNUSMY06,BolEnd09}, and on many-particle quantum 
graphs see Refs.~\citenum{BKSingular,BKContact}.

A definition of BEC that also applies to a system of interacting bosons was 
given in Ref.~\citenum{PO56} and employs the reduced one-particle 
density matrix obtained from the canonical density matrix (at inverse 
temperature $\beta=\frac{1}{T}$),
\begin{equation}\label{fulldensitymatrix}
\rho_{N}(x,y)=\frac{1}{Z_{N}(\beta)}\sum_{n} e^{-\beta E^{N}_{n}}\bar\Psi_{n}(x)
\Psi_{n}(y)\ ,
\end{equation}
by tracing out all degrees of freedom except one. Here $\Psi_{n}(x)$ is the 
$n$-th eigenfunction of the $N$-particle system with eigenvalues $E^{N}_{n}$, 
and $Z_{N}(\beta)=\sum_{n}e^{-\beta E^{N}_{n}}$ is the canonical partition 
function. Condensation is defined to occur when the largest eigenvalue 
of the reduced density matrix is asymptotically of order one as
$N \rightarrow \infty$. Note that this limit (the thermodynamic limit) is 
accompanied by the limit $V \rightarrow \infty$, where $V$ is the volume of 
the one-particle configuration space, such that the particle density remains 
fixed.

Unfortunately, it is in general very difficult to prove (or disprove) BEC in
an interacting system in the sense of Penrose-Onsager. Instead we shall employ
the connection between BEC and phase transitions in the case of interacting 
bosons on a graph.

The classical configuration space for a particle on a graph is a compact
metric graph, i.e., a finite, connected graph 
$\Gamma = (\mathcal{V},\mathcal{E})$ with vertex set 
$\mathcal{V} = \{v_1,\dots,v_{V}\}$ and edge set 
$\mathcal{E}=\{e_1,\dots,e_{E}\}$. The edges are identified with intervals 
$[0,l_e]$, $e=1,\dots,E$, thus assigning lengths to intervals. This then 
introduces a metric on the graph. Note that a graph is called \textit{compact} 
when all lengths are finite.

For the one-particle quantum system the Hilbert space is 
\begin{equation}
\mathcal{H}_1=L^{2}(\Gamma)=\bigoplus_{e=1}^{E}L^{2}(0,l_{e}).
\end{equation}
The quantum dynamics shall be generated by a self-adjoint realisation of the
Laplacian $-\Delta_{1}$. As a differential expression this acts on each 
edge-component $f_e$ of a function $F=(f_1,\dots,f_E)\in C^{\infty}_{0}(\Gamma)$ 
as the negative second derivative, $-\Delta_1 F=(-f_1'',\dots,-f_E'')$. As 
shown in Ref.~\citenum{Kuc04}, one way to arrive at all possible self-adjoint
realisations of the Laplacian uses quadratic forms, 
\begin{equation}
\label{quadform1}
 Q_{1}[F] = \sum_{e=1}^{E}\int_0^{l_e}|f^{\prime}(x)|^2 \mathrm{d} x -
 \langle F_{bv},L_1 F_{bv}\rangle_{\mathbb{C}^{2E}}\ ,
\end{equation}
with domain 
\begin{equation}
 \mathcal{D}_{Q_1} = \{F\in H^1(\Gamma); P_1 F_{bv}=0 \}\ .
\end{equation} 
Here $F_{bv} = (f_1(0),\dots,f_E(0),f_1(l_1),\dots,f_E(l_E))^{T}\in\mathbb{C}^{2E}$
is the vector of boundary values, $P_1$ is an orthogonal projection on 
$\mathbb{C}^{2E}$ and $L_1$ a self-adjoint endomorphism of $\ker{P_1}$.

The $N$-particle Hilbert space is the $N$-fold tensor product 
$\mathcal{H}_N=\mathcal{H}_1 \otimes\cdots\otimes \mathcal{H}_1$,
\begin{equation}
\label{NHilbertSpace}
 \mathcal{H}_N =L^{2}(\Gamma_{N})=\bigoplus_{e_{1}e_{2}...e_{N}}
 L^{2}(D_{e_{1}e_{2}...e_{N}}),
\end{equation}
where $D_{e_{1}e_{2}...e_{N}}=(0,l_{e_{1}})\times\cdots\times (0,l_{e_{N}})$ are 
$N$-dimensional hypercubes. In order to introduce contact interactions
that are formally given by a Hamiltonian
\begin{equation}
\label{deltainter}
 H_N =-\Delta_N + \alpha\sum_{i\neq j}\delta (x_{e_i}-x_{e_j})
\end{equation}
we also need to dissect the $N$-particle configuration space further by
cutting the hypercubes along all hypersurplanes defined by $x_{e_{i}}=x_{e_{j}}$ 
when $e_{i}=e_{j}$. We denote \eqref{NHilbertSpace} as 
$L^2(\Gamma^{\ast}_{N})$ when taking this further dissection into account.
The bosonic subspace is denoted as $L^{2}_{B}(\Gamma^{\ast}_{N})$.

As in the one-particle case, we require the $N$-particle Hamiltonian to be a 
self-adjoint realisation of the Laplacian $-\Delta_N$. As a differential 
expression it acts as
\begin{equation}
 (-\Delta_{N}\Psi)_{e_{1}\dots e_{N}} =
  -\left(\frac{\partial^{2}}{\partial x_{e_{1}}^{2}}+\cdots
  +\frac{\partial^{2}}{\partial x_{e_{N}}^{2}}\right)\psi_{e_{1}\dots e_{N}}\ ,
\end{equation}
on functions $\Psi\in C^{\infty}_{0}(\Gamma_{N})$. Self-adjoint (bosonic)
realisations of $-\Delta_{N}$ are obtained through suitable quadratic forms. 
For this, we first define vectors of boundary values 
\begin{equation}
\label{BVI}
 \Psi_{bv}(\boldsymbol{y}) = \begin{pmatrix} \sqrt{l_{e_{2}}\dots l_{e_{N}}} 
 \psi_{e_{1}\dots e_{N}}(0,l_{e_{2}}y_{1},\dots,l_{e_{N}}y_{N-1}) \\ 
 \sqrt{l_{e_{2}}\dots l_{e_{N}}} 
 \psi_{e_{1}\dots e_{N}}(l_{e_{1}},l_{e_{2}}y_{1},\dots,l_{e_{N}}y_{N-1}) \end{pmatrix},
\end{equation}
with $\boldsymbol{y} \in [0,1]^{N-1}$. The desired quadratic form now reads
\begin{equation}
\label{QuadFormHard}
\begin{split}
 Q^{(N)}_{B}[\Psi] 
  &= N \sum_{e_{1}\dots e_{N}}\int_{0}^{l_{e_{1}}}\dots\int_{0}^{l_{e_{N}}} 
    |\psi_{e_{1}\dots e_{N},x_{e_1}}(x_{e_{1}},\dots,x_{e_{N}})|^{2}\ \mathrm{d} x_{e_N}\dots
    \mathrm{d} x_{e_1} \\
  &\quad -N\int_{[0,1]^{N-1}}\langle\Psi_{bv},L_{N}(\boldsymbol{y})\Psi_{bv} 
    \rangle_{\mathbb{C}^{2E^{N}}} \mathrm{d}\boldsymbol{y},
\end{split}
\end{equation}
and is defined on
\begin{equation}
\label{QNformharddomain}
 \mathcal{D}_{Q^{(N)}_{B}} = \{\Psi \in H^{1}_{0,B}(\Gamma^{\ast}_{N});\ 
 P_{N}(\boldsymbol{y})\Psi_{bv}(\boldsymbol{y})=0\ \text{for a.e.}\ 
 \boldsymbol{y}\in [0,1]^{N-1}\}.
\end{equation}
Here $\Psi\in H^{1}_{0,B}(\Gamma^{\ast}_{N})\subset H^{1}_{B}(\Gamma^{\ast}_{N})$ if 
each component $\psi_{e_{1}e_{2}...e_{N}}$ is in $H^1$ and vanishes on the 
hyperplanes $x_{e_{i}}=x_{e_{j}}$. Moreover, 
$P_N,L_N:[0,1]^{N-1} \rightarrow M(2E^{N},\mathbb{C})$ are bounded and 
measurable maps such that $P_{N}(\boldsymbol{y})$ is an orthogonal projection 
and $L_{N}(\boldsymbol{y})$ is a self-adjoint endomorphism of 
$\ker{P_{N}(\boldsymbol{y})}$. 

The self-adjoint realisation of $-\Delta_{N}$ associated with 
$ Q^{(N)}_{B}[\cdot]$ is denoted as 
$(-\Delta_{N},\mathcal{D}^{\infty}_{N}(P_{N},L_{N}))$, see 
Ref.~\citenum{BKContact}. Due to the Dirichlet 
conditions along the hyperplanes $x_{e_{i}}=x_{e_{j}}$ this operator is a rigorous 
version of the Hamiltonian \eqref{deltainter} in the limit $\alpha\to\infty$.
Therefore it represents \textit{hardcore interactions}. It is important to note 
that the coordinate dependence of the maps $P_{N}$ and $L_{N}$ leads to 
(additional) singular many-particle interactions that are localised in the 
vertices of the graph.\cite{BKSingular} 
\section{BEC in non-interacting systems many-particle}
In this section we provide a complete classification of non-interacting 
many-particle systems on general compact quantum graph in terms of the
presence or absence of BEC. Since non-interacting Hamiltonians 
are entirely determined by 
corresponding one-particle Hamiltonians $(-\Delta_{1},\mathcal{D}_{1}(P_1,L_1))$, 
it is sufficient to refer to the latter only.
\begin{definition} 
Let $\Gamma$ be a compact, metric graph with edge lengths $l_1,...,l_{E}$. 
The thermodynamic limit (TL) consists of the scaling $l_{e}\mapsto\eta l_{e}$ 
and taking the limit $\eta\rightarrow\infty$.
\end{definition}
As a first result we obtain the following.\cite{BKBEC}
\begin{lemma} 
If the one-particle Laplacian $(-\Delta_{1},\mathcal{D}_{1}(P_1,L_1))$ is such 
that $L_1$ is negative semi-definite, no 
BEC occurs in the corresponding free Bose gas at finite temperature ($T > 0$) in the thermodynamic limit.
\end{lemma}
The proof uses the formalism of the grand-canonical ensemble and, via a
bracketing construction, compares the given systems with free Bose gases
with Dirichlet- and Neumann boundary conditions in the vertices.
The number of particles is not fixed and the particle density is adjusted via 
the chemical potential $\mu$. 

Our main result in this section is now the following.\cite{BKBEC}
\begin{theorem}\label{MainTheoremNON}
Let a free Bose gas be given on a quantum graph with a one-particle 
Laplacian $(-\Delta_{1},\mathcal{D}_{1}(P_1,L_1))$ such that $L_1$ has at least 
one positive eigenvalue. Then, in the thermodynamical limit, there is a 
critical temperature $T_c>0$ such that BEC occurs below $T_c$. 
\end{theorem}
In order to prove this statement one shows that the one-particle ground state 
energy (which is negative since $L_1$ has a positive eigenvalue, compare
\eqref{quadform1}) remains negative in the thermodynamic limit. Indeed, one 
uses the lower bound for the one-particle ground state energy of 
Ref.~\citenum{KS06} as well as a suitable Rayleigh quotient to prove that
the ground state energy converges to $-L^{2}_{max}$, 
where $L_{max} > 0$ is the largest positive eigenvalue of $L_1$. One also uses 
that the number of negative eigenstates is bounded from above~\cite{KS06} (even
in the thermodynamic limit) and applies the trace formula for quantum
graphs\cite{BolEnd09}. With standard arguments the Theorem then follows.
\section{Interacting many-particle systems}
In this final section we consider systems of bosons interacting via singular 
interactions localised at the vertices of the graph as well as hardcore 
contact interactions, i.e., we consider (bosonic) self-adjoint realisations 
$(-\Delta_{N},\mathcal{D}^{\infty}_{N}(P_{N},L_{N}))$ of the $N$-particle 
Laplacian. Our goal is to prove that no phase transitions (in the free energy 
density) are present, indicating an absence of condensation. For this 
recall that the free energy density of a sequence of $N$-particle Hamiltonians 
$H_N$ with discrete spectra $\{E^{N}_{n}\}$ is defined as
\begin{equation}
 f(\beta,\mu)=-\lim_{V \rightarrow \infty}\frac{1}{\beta V}
 \log{\sum_{N=1}^{\infty}e^{N\beta\mu}\tr_{\mathcal{H}_{N}}}e^{-\beta H_N}\ .
\end{equation}
The vanishing of functions in the domain $\mathcal{D}^{\infty}_{N}(P_{N},L_{N})$ 
along the hyperplanes $x_{e_{i}}=x_{e_{j}}$ allows to define a Fermi-Bose mapping,
relating the bosonic free-energy density to a fermionic one (which is known
explicitly). This leads to the following result.\cite{BKBEC}
\begin{theorem}\label{MainTheoremInteracting}
Let $(-\Delta_{N},\mathcal{D}^{\infty}_{N}(P_{N},L_{N}))_{N \in \mathbb{N}}$ be a 
family of bosonic Laplacians with repulsive hardcore interactions, indexed by 
the particle number $N$. Assume that for this family the operator of 
multiplication with $L_N(\cdot)$ on $L^2([0,1]^{N-1})$ is uniformly bounded
with respect to $N$. Then the bosonic, grand-canonical, free-energy 
density $f(\beta,\mu)$ is given by
\begin{equation}
 f(\beta,\mu) = -\frac{1}{\pi \beta}\int_{0}^{\infty}\log{\left(
 1+e^{-\beta(k^{2}-\mu)}\right)}\ \mathrm{d} k\ .
\end{equation}
This function is smooth and, hence, there exists no phase transition.
\end{theorem}
It is important to note that Theorem \ref{MainTheoremInteracting} holds 
\textit{independently} of the singular interactions in the vertices, i.e.,
independently of the maps $P_N$ and $L_N$. Hence, even when BEC occurs without 
the hardcore interactions, switching on the latter destroys the
condensation.
\bibliographystyle{ws-procs975x65}
\bibliography{LiteratureQMATH13}

\end{document}